\documentclass[twocolumn,prb,amssymb]{revtex4}

\newcommand{\be}{\begin{equation}}
\newcommand{\ee}{\end{equation}}

\def\Rbol{{\stackrel{\circ}{R}}{}}

\def\Abol{{\stackrel{~\circ}{A}}{}}

{}
{}

\begin{document}

\title{Gravitation and Duality Symmetry}

\author{V. C. de Andrade}
\email{andrade@fis.unb.br}
\affiliation{Instituto de F\'{\i}sica, Universidade de Bras\'\i lia \\ 70919-970
Bras\'\i lia DF, Brazil}
\author{A. L. Barbosa}
\email{analucia@ift.unesp.br}
\author{J. G. Pereira}
\email{jpereira@ift.unesp.br}
\affiliation{Instituto de F\'{\i}sica Te\'orica, Universidade Estadual Paulista \\ 
Rua Pamplona 145, 01405-900 S\~ao Paulo, Brazil}

\begin{abstract}
By generalizing the Hodge dual operator to the case of soldered bundles, and working in the
context of the teleparallel equivalent of general relativity, an analysis of the duality
symmetry in gravitation is performed. Although the basic conclusion is that, at least in the
general case, gravitation is not dual symmetric, there is a particular theory in which this
symmetry shows up. It is a self dual (or anti-self dual) teleparallel gravity in which, due to
the fact that it does not contribute to the interaction of fermions with gravitation, the
purely tensor part of torsion is assumed to vanish. The ensuing fermionic gravitational
interaction is found to be chiral. Since duality is intimately related to renormalizability,
this theory may eventually be more amenable to renormalization than teleparallel gravity or
general relativity.
\end{abstract}

\maketitle

\section{Introduction}

Duality symmetry is an important property of sourceless gauge theories. When it holds, the field
equation of the theory is just the Bianchi identity written for the dual of the field strength.
This means essentially that, if we know the geometrical background, that is, the Bianchi
identity, we know the dynamics, or the field equations. In this sense, it is usual to assert
that gauge theories are fundamentally geometric: the dynamical field equations are related by
duality to the purely geometric Bianchi identities.

Gravitation, on the other hand, as described by general relativity, is not dual symmetric. In
fact, when written for the dual of the curvature tensor, its Bianchi identities do not yield
Einstein's equation. Even though, duality symmetry in gravitation has attracted considerable
attention.\cite{saco} In particular, it has been shown recently that duality invariance holds
for linearized general relativity, that is, for free massless spin-2 fields,\cite{ht04} a
result that has also been extended to all free bosonic and fermionic massless gauge fields in
flat four-dimensional spacetime.\cite{ds04} Now, as a gauge theory\cite{pr} for the translation
group,\cite{footnote1} the teleparallel equivalent of general relativity,\cite{tp} or
teleparallel gravity for short, despite equivalent to general relativity, presents several
peculiar properties which distinguish it from general relativity. For example, similarly to
Maxwell's theory, it describes the gravitational interaction, not through a geometrization of
spacetime, but by a true force.\cite{anpe1} As a consequence of this property, teleparallel
gravity is found not to require the weak equivalence principle to describe the gravitational
interaction,\cite{wep} although it can comply with its seemingly validity.\cite{exp} Another
important point refers to the Lagrangian of the gravitational field. In general relativity, as
is well known, it is linear in the curvature, whereas in teleparallel gravity it is quadratic
in the torsion tensor, the field strength of the theory, a typical characteristic of {\it
internal} (or Yang--Mills type) gauge theories.

Now, a fundamental difference between internal and external gauge theories is the presence in
the latter of a solder form,\cite{konu} whose components constitute the tetrad field. The
presence of this form gives rise to new kind of contractions, which are not possible in
internal gauge theories. As a consequence, the gauge Lagrangian, as well as the field equation,
will include additional terms in relation to the usual Yang-Mills theories. These additional
terms, as we are going to see, can be taken into account through a generalization for soldered
bundles of the concept of {\it dual}. For non-soldered bundles, of course, it might reduce to
the usual definition of dual. Relying then on the gauge structure of teleparallel gravity,
which makes it closer to the usual Yang-Mills gauge theories, and using the generalized dual
definition alluded to above, the basic purpose of this paper will be to analyze whether
teleparallel gravity, under some specific conditions, may eventually result dual symmetric, a
typical property of internal gauge theories. We begin in the next section with a review of the
strictly necessary ingredients of the teleparallel equivalent of general relativity.

\section{Fundamentals of Teleparallel Gravity}

According to the gauge structure of teleparallel gravity, to each point of spacetime there is
attached a Minkowski tangent space, on which the translation (gauge) group acts. We use the
Greek alphabet $\mu, \nu, \rho, \dots = 0, 1, 2, 3$ to denote spacetime indices and the Latin
alphabet $a, b, c, \dots = 0, 1, 2, 3$ to denote algebraic indices related to the tangent
Minkowski spaces, whose metric is chosen to be $\eta_{a b} = {\rm diag} (+1, -1, -1, -1)$. As
a gauge theory for translations, the fundamental field of teleparallel gravity is the
translational gauge potential $B^a{}_\mu$, a 1-form assuming values in the Lie algebra of the
translation group\cite{anpe1}
\be
B_\mu = B^a{}_\mu \, P_a,
\ee
with $P_a = \partial_a$ the generators of infinitesimal translations. Under a local translation
of the tangent space coordinates $\delta x^a = \epsilon^a(x)$, the gauge potential transforms
according to
\be
B^{\prime a}{}_\mu = B^a{}_\mu - \partial_\mu \epsilon^a.
\label{btrans}
\ee
It appears naturally as the nontrivial part of the tetrad field $h^{a}{}_{\mu}$:
\be
h^a{}_\mu = \partial_\mu x^a + B^a{}_\mu.
\label{tetrada}
\ee
Notice that, whereas the tangent space indices are raised and lowered with the Minkowski metric
$\eta_{a b}$, the spacetime indices are raised and lowered with the spacetime metric
\be
g_{\mu \nu} = \eta_{a b} \; h^a{}_\mu \; h^b{}_\nu.
\label{gmn}
\ee

The above tetrad can be used to construct the so called Weit\-zen\-b\"ock connection
\begin{equation}
\Gamma^{\rho}{}_{\mu\nu} = h_{a}{}^{\rho}\partial_{\nu}h^{a}{}_{\mu},
\label{carco}
\end{equation}
which introduces the distant parallelism in the four-dimensional spacetime manifold. It
is a connection presenting torsion, but no curvature. Its torsion,
\begin{equation}
T^{\rho}{}_{\mu\nu} = \Gamma^{\rho}{}_{\nu\mu} - 
\Gamma^{\rho}{}_{\mu\nu},
\label{tor}
\end{equation}
is related to the translational gauge field strength $F^a{}_{\mu \nu}$ by
\begin{equation}
F^a{}_{\mu \nu} \equiv \partial_\mu B^a{}_{\nu} - \partial_\nu B^a{}_{\mu} =
h^a{}_\rho \; T^\rho{}_{\mu \nu}.
\label{gfs}
\end{equation}
The Weitzenb\"ock connection is related to the Levi-Civita connection
${\stackrel{\circ}{\Gamma}}{}^{\rho}{}_{\mu\nu}$ of the spacetime metric $g_{\mu\nu}$ through
\begin{equation}
\Gamma^{\rho}{}_{\mu\nu} = {\stackrel{\circ}{\Gamma}}{}^{\rho}{}_{\mu\nu} 
+ K^{\rho}{}_{\mu\nu},
\label{rela}
\end{equation}
where
\begin{equation}
K^{\rho}{}_{\mu \nu} = \textstyle{\frac{1}{2}} \left( 
T_{\mu}{}^{\rho}{}_{\nu} + T_{\nu}{}^{\rho}{}_{\mu} 
- T^{\rho}{}_{\mu \nu} \right)
\label{contorsion}
\end{equation}
is the contortion tensor. It is important to remark that curvature and torsion are properties
of a connection, not of spacetime.\cite{livro} Notice, for example, that the Christoffel and
the Weitzenb\"ock connections, which have different curvature and torsion tensors, are defined
on the very same spacetime manifold.

The Lagrangian of the teleparallel equivalent of general relativity is\cite{anpe1}
\be
{\cal L} = \frac{h}{8 k^2} \left[
T^\rho{}_{\mu \nu} T_\rho{}^{\mu \nu} + 2 \,
T^\rho{}_{\mu \nu} T^{\nu \mu} {}_\rho - 4 \, T_{\rho \mu}{}^{\rho}
T^{\nu \mu}{}_\nu \right].
\label{lagr3}
\ee
where $k^2 = 8 \pi G/c^{4}$ and $h = {\rm det}(h^{a}{}_{\mu})$. The first term corresponds
to the usual Lagrangian of gauge theories. In the gravitational case, however, owing to the
presence of a tetrad field, which are components of the solder form, algebra and spacetime
indices can now be changed into each other, and in consequence new contractions turn out to be
possible. It is exactly this possibility that gives rise to the other two terms of the above
Lagrangian. If we define the tensor
\begin{equation}
S^{\rho\mu\nu} = - S^{\rho\nu\mu} =
\left[ K^{\mu\nu\rho} - g^{\rho\nu}\,T^{\sigma\mu}{}_{\sigma} 
+ g^{\rho\mu}\,T^{\sigma\nu}{}_{\sigma} \right],
\label{S}
\end{equation}
usually called superpotential, it can be rewritten in the form\cite{maluf}
\begin{equation}
{\mathcal L} =
\frac{ h}{4 k^2} \; T^\rho{}_{\mu\nu} \, S_\rho{}^{\mu\nu}.
\label{gala}
\end{equation}

Performing a variation with respect to the gauge potential, we find the teleparallel version
of the gravitational field equation,
\begin{equation}
\partial_\sigma(h S_\lambda{}^{\rho \sigma}) -
k^2 \, (h t_\lambda{}^\rho) = 0,
\label{eqs1}
\end{equation}
where
\begin{equation}
h \, t_\lambda{}^\rho = \frac{h}{k^2} \left( \Gamma^\mu{}_{\nu\lambda} \,
S_{\mu}{}^{\rho \nu} - \frac{1}{4} \, \delta_\lambda{}^\rho \,
T^\theta{}_{\mu\nu} \, S_\theta{}^{\mu\nu} \right)
\label{emt1}
\end{equation}
is the energy-momentum pseudotensor of the gravitational field. This is actually the Noether
current, which is conserved when the field equation (\ref{eqs1}) is satisfied.\cite{kopo} If
we write
\be
S_a{}^{\rho \sigma} = h_a{}^\lambda \, S_\lambda{}^{\rho \sigma},
\label{ShS}
\ee
the above field equation acquires the form
\begin{equation}
\partial_\sigma(h S_a{}^{\rho \sigma}) -
k^2 \, (h j_a{}^\rho) = 0,
\label{eqs1bis}
\end{equation}
where\cite{prl}
\begin{equation}
j_a{}^\rho \equiv - \frac{\partial {\mathcal L}}{\partial h^a{}_\rho} =
\frac{h_a{}^\lambda}{k^2}
\left( F^c{}_{\mu \lambda} \, S_c{}^{\mu \rho} - \frac{1}{4} \, \delta_\lambda{}^\rho \,
F^c{}_{\mu \nu} \, S_c{}^{\mu \nu} \right).
\label{emt1bis}
\end{equation}
represents the tensorial form of the gravitational energy-momentum current, whose form is
analogous to the energy-momentum tensor of the Yang-Mills field.\cite{iz}

\section{Generalized Dual Operation for Soldered Bundles}

The possibility of contracting internal with external indices is a typical property of gauge
theories for gravitation. Technically, it is revealed by the presence of a solder form, whose
components constitute the tetrad field. This property gives rise to deep changes in relation to
the usual internal, that is, non-soldered gauge theories. A crucial difference, as we have
already seen in the case of teleparallel gravity, is the appearance of new terms in the gauge
Lagrangian. Now, the existence of these new terms can be attributed to a new definition of
dual tensor, which turns up as a natural generalization for soldered bundles. The fundamental
point is to require that, like in any other gauge theory, the action functional of teleparallel
gravity be of the form
\begin{equation}
{\mathcal S} =  \frac{1}{k^2} \int \, \mbox{tr} (F \wedge {^\ast}F),
\label{action1}
\end{equation}
or equivalently,
\begin{equation}
{\mathcal S} = \frac{1}{k^2} \int \, \eta_{ab} \, F^a \wedge {^\ast}F^b,
\label{action2}
\end{equation}
where
\begin{equation}
F^a = \textstyle{\frac{1}{2}} \, F^a{}_{\mu\nu} \, dx^\mu \wedge dx^\nu
\label{tform}
\end{equation}
is the torsion 2-form, and
\begin{equation}
{^\ast}F^a = \textstyle{\frac{1}{2}} \, ^*F^a{}_{\rho \sigma} \, dx^\rho \wedge
dx^\sigma 
\label{dualtform}
\end{equation}
is the corresponding dual form. Using Eqs.~(\ref{tform}) and (\ref{dualtform}), the
action functional (\ref{action2}) becomes
\begin{equation}
{\mathcal S} =
\frac{1}{4 k^2} \int \, \eta_{ab} \, F^a_{\mu\nu} \;
^*{F}^b{}_{\rho \sigma} \, dx^\mu \wedge dx^\nu \wedge dx^\rho \wedge dx^\sigma.
\label{action3}
\end{equation}
Using the identity
\begin{equation}
dx^\mu \wedge dx^\nu \wedge dx^\rho \wedge dx^\sigma = -
\epsilon^{\mu \nu \rho \sigma} \, h \, d^4x,
\end{equation}
where $\epsilon^{\mu \nu \rho \sigma}$ is the totally anti-symmetric Levi-Civita tensor, with
$\epsilon_{0123} = h$, as well as the relations (\ref{gmn}) and (\ref{gfs}), the action
functional can be rewritten in the form
\begin{equation}
{\mathcal S} = -
\frac{1}{4 k^2} \int \, T_{\alpha\mu\nu} \;
^*T^{\alpha}{}_{\rho \sigma} \, \epsilon^{\mu \nu \rho \sigma} \; h \, d^4x.
\label{action4}
\end{equation}
If we define a generalized dual torsion according to
\begin{equation}
^*{T}^\alpha{}_{\rho \sigma} = \textstyle{\frac{1}{2}} \, \epsilon_{\theta \lambda \rho
\sigma} \, S^{\alpha \theta \lambda}, 
\label{soldual}
\end{equation}
we obtain
\begin{equation}
{\mathcal S} =
\frac{1}{4 k^2} \int h \ T_{\rho\mu\nu} \, S^{\rho\mu\nu} \; d^4x,
\label{action5}
\end{equation}
which yields the teleparallel gauge Lagrangian (\ref{gala}). We see in this way that the dual
definition (\ref{soldual}) automatically takes into account all possible index contractions,
and can accordingly be considered as a generalization of the dual operation for soldered
bundles. In fact, when rewritten in terms of the torsion tensor, the dual operation
(\ref{soldual}) becomes
\begin{equation}
^*{T}^\alpha{}_{\rho \sigma} = \textstyle{\frac{1}{4}} \, \epsilon_{\mu \nu \rho \sigma}
\, \left(T^{\alpha \mu \nu} + 2 T^{\mu \alpha \nu} - 4 g^{\alpha \nu} \, T^{\theta \mu}{}_\theta
\right).
\label{soldualbis}
\end{equation}
Up to a numerical factor, the first term corresponds to the usual definition of dual for
non-soldered bundles. In the case of soldered bundles, however, owing to the presence of a
tetrad field, algebra and spacetime indices can be changed into each other, and in consequence
new contractions show up in the dual operation. These new possibilities are represented by the
second and third terms of Eq.~(\ref{soldualbis}).

It is important to remark that the definition (\ref{soldual}) presents all properties to be
considered as a consistent dual definition. For example, when applied twice it gives back, up to
a sign, the torsion 2-form:
\begin{equation}
{^{\ast \ast}}F^a = - F^a.
\end{equation}
In components, this relation reads
\begin{equation}
{^{\ast \ast}}T^\alpha{}_{\rho \sigma} \equiv
\textstyle{\frac{1}{2}} \, \epsilon_{\mu \nu \rho \sigma} \, ^*{S}^{\alpha \mu \nu} =
- T^\alpha{}_{\rho \sigma},
\label{dualdual}
\end{equation}
which is identical to the dual property of the electromagnetic field strength in a
four-dimensional spacetime with signature $s=2$. This is a solid property of both
electromagnetism and gravitation; its violation would require either a change in the dimension
of spacetime or in the signature of the metric.

\section{Duality Symmetry in Gravitation}

Let us consider now the first Bianchi identity of teleparallel gravity, which is given
by\cite{bianchi}
\begin{equation}
\partial_\rho F^a{}_{\mu \nu} + \partial_\nu F^a{}_{\rho \mu} +
\partial_\mu F^a{}_{\nu \rho} = 0,
\label{bid}
\end{equation}
or equivalently
\begin{equation}
\epsilon^{\lambda \rho \mu \nu} \; \partial_\rho F^a{}_{\mu \nu} = 0.
\label{bid2}
\end{equation}
Written for the dual, it reads
\begin{equation}
\epsilon^{\lambda \rho \mu \nu} \; \partial_\rho (^*{F}^a{}_{\mu \nu}) = 0.
\label{bidbis}
\end{equation}
Using the property\cite{wald}
\begin{equation}
\partial_\rho (h \epsilon^{\lambda \rho \mu \nu}) = 0,
\label{lccoder}
\end{equation}
as well as the dual definition (\ref{soldual}), it can be written in the form
\be
\partial_\sigma(h S_a{}^{\rho \sigma}) = 0,
\label{bifordual}
\ee
from where we see that the Bianchi identity written for the generalized dual torsion
yields a vanishing tensorial energy-momentum current:
\be
j_a{}^\rho = 0.
\ee
Using the identification (\ref{ShS}), and also the definition (\ref{carco}) of the
Weitzenb\"ock connection, equation (\ref{bifordual}) can be written in the equivalent
form
\begin{equation}
\partial_\sigma(h S_\lambda{}^{\rho \sigma}) - h \, \Gamma^\mu{}_{\nu \lambda} \,
S_\mu{}^{\rho \nu} + h \, T^\mu{}_{\nu \lambda} \, S_\mu{}^{\nu \rho} = 0,
\label{eqs2}
\end{equation}
which, compared with the field equation (\ref{eqs1}), yields a pseudocurrent of the form
\be
h t_\lambda{}^\rho =  \frac{h}{k^2} \left( \Gamma^\mu{}_{\nu\lambda} \,
S_{\mu}{}^{\rho \nu} - T^\mu{}_{\nu \lambda} \, S_\mu{}^{\nu \rho} \right).
\ee
Notice that, although the tensorial current $j_a{}^\rho$ vanishes, the corresponding
pseudotensor $t_\lambda{}^\rho$ will never be zero.

Comparing the above equations---obtained from the Bian\-chi identity written for the
generalized dual tor\-sion---with the corresponding field equations of tele\-parallel gravity,
we see that the condition for the theory to present duality symmetry is that
\be
T^\mu{}_{\nu \lambda} \, S_\mu{}^{\nu \rho} = 
\textstyle{\frac{1}{4}} \, \delta_\lambda{}^\rho \, T^\mu{}_{\nu \theta} \, S_\mu{}^{\nu
\theta}.
\label{condual}
\ee
In fact, in this case Eq.~(\ref{eqs2}) coincides with the field equation (\ref{eqs1}),
with the canonical energy-momentum pseudotensor given by (\ref{emt1}). This is, however, a
quite restrictive condition, which seems not to be realized, at least in the general case we
have considered here. However, as we are going to see, under some specific conditions,
gravitation may become dual symmetric.

\section{In Search of a Dual Symmetric Gravitation}

\subsection{Torsion Decomposition and Spinors}

As is well known, torsion can be decomposed in irreducible components under the global
Lorentz group:\cite{hb73}
\be
T_{\lambda \mu \nu} = \textstyle{\frac{2}{3}} \left(t_{\lambda \mu \nu} -
t_{\lambda \nu \mu} \right) + \frac{1}{3} \left(g_{\lambda \mu} v_\nu -
g_{\lambda \nu} v_\mu \right) + \epsilon_{\lambda \mu \nu \rho} \, a^\rho.
\label{deco}
\ee
In this expression, $v_\mu$ and $a^\rho$ represent the vector and axial parts of torsion,
defined respectively by
\be
v_{\mu} =  T^{\nu}{}_{\nu \mu}
\label{pt2}
\ee
and
\be
a^{\mu} = \textstyle{\frac{1}{6}} \epsilon^{\mu\nu\rho\sigma} \, T_{\nu\rho\sigma},
\label{pt3}
\ee
whereas $t_{\lambda \mu \nu}$ is the purely tensor part, given by
\be
t_{\lambda \mu \nu} = \textstyle{\frac{1}{2}} \left(T_{\lambda \mu \nu} +
T_{\mu\lambda \nu} \right) + \frac{1}{6} \left(g_{\nu \lambda} v_\mu +
g_{\nu \mu} v_\lambda \right) - \frac{1}{3} g_{\lambda \mu} \, v_\nu.
\label{pt1}
\ee
Now, in teleparallel gravity, the gravitational interaction of a Dirac spinor is
well known to involve only $v_\mu$ and $a_\mu$. In fact, the teleparallel Dirac equation
reads\cite{mospe}
\begin{equation}
i \, \hbar \, \gamma^{\mu} \left(
\partial_{\mu} - \frac{1}{2} \, v_{\mu} - \frac{3 i}{4} \, a_{\mu} \gamma^{5}
\right)  \psi = m c \, \psi,
\label{teledirac}
\end{equation}
where $\gamma^{\mu} = \gamma^a h_a{}^\mu$, with $\gamma^a$ the Dirac matrices, and
$\gamma^{5}=\gamma_{5} =i \gamma^{0}\gamma^{1}\gamma^{2}\gamma^{3}$. This means essentially
that the purely tensor piece $t_{\lambda \mu \nu}$ is irrelevant for the description of the
gravitational interaction of spinor fields. Relying on this property, we consider then the case
in which $t_{\lambda \mu \nu}$ vanishes, and torsion acquires the form 
\be
T_{\lambda \mu \nu} = \textstyle{\frac{1}{3}} \left(g_{\lambda \mu} v_\nu -
g_{\lambda \nu} v_\mu \right) + \epsilon_{\lambda \mu \nu \rho} \, a^\rho.
\label{microtor}
\ee
The corresponding contortion tensor is
\begin{equation}
K^{\rho \mu \nu} = \textstyle{\frac{1}{3}} \left(g^{\nu \rho} v^\mu -
g^{\nu \mu} v^\rho \right) - \frac{1}{2} \epsilon^{\nu \rho \mu \lambda} \, a_\lambda,
\label{microcontor}
\end{equation}
whereas the superpotential $S_\rho{}^{\mu\nu}$ becomes
\be
S^{\rho \mu \nu} = - \textstyle{\frac{2}{3}} \left(g^{\rho \mu} v^\nu -
g^{\rho \nu} v^\mu \right) - \frac{1}{2} \epsilon^{\rho \mu \nu \lambda} \, a_\lambda.
\label{microS}
\ee
The gravitational Lagrangian (\ref{gala}), in consequence, turns out to be
\be
{\mathcal L} =
\frac{h}{2 k^2} \left( - \frac{2}{3} v_\mu v^\mu +
\frac{3}{2} a_\mu a^\mu \right).
\ee

Now, using the above expressions, it can be easily verified that
\begin{eqnarray}
T_{\mu \nu \lambda} \, S^{\mu \nu \rho} = - \left(\textstyle{\frac{4}{9}} v_\lambda v^\rho
+ \frac{1}{2} \epsilon_\lambda{}^{\rho \gamma \mu} a_\gamma v_\mu +
a_\lambda a^\rho \right) \nonumber \\
- \delta_\lambda{}^\rho \left(\textstyle{\frac{2}{9}} v_\mu v^\mu -
a_\mu a^\mu \right)
\label{c1}
\end{eqnarray}
and
\be
T_{\mu \nu \theta} \, S^{\mu \nu \theta} = - \textstyle{\frac{4}{3}} v_\mu v^\mu +
3 a_\mu a^\mu.
\label{c2}
\ee
From these expressions we see that condition (\ref{condual}) is not satisfied, and
consequently, even in the case considered here, characterized by the vanishing of $t_{\lambda
\mu \nu}$, gravitation is not dual symmetric.

\subsection{Self Dual and Anti-Self Dual Fields}

Duality symmetry plays a key role in the quantization of non-linear.\cite{ash1} Furthermore,
significant simplifications arise in the case of self dual and anti-self dual
fields.\cite{penrose1} In fact, in the classical theory, Yang--Mills simplifies enormously in
the self dual (or anti-self dual) sector. In particular, the field equations become quite
simple and transparent. For these reasons, it is tempting to consider these cases, mainly when
dealing with the quantization problem.\cite{ash2} This is what we are going to do next in the
case of teleparallel gravity.

To begin with, we note that the dual definition (\ref{soldual}) implies the following
properties for the axial and vector torsions:
\be
^*{a}_\mu = - \textstyle{\frac{2}{3}} \; v_\mu \quad \mbox{and} \quad
^*{v}_\mu = \textstyle{\frac{3}{2}} \; a_\mu.
\label{dualp1}
\ee
We see clearly that they are neither self dual nor anti-self dual. In spite of this, by
appropriately redefining the spin connection, it is always possible to consider the self dual
and anti-self dual sectors of the theory.\cite{ash3} However, we prefer here to follow a more
stringent alternative, which consists in requiring that torsion be self dual (or anti-self
dual). In other words, we require that
\be
^*{a}_\mu = \pm i a_\mu \quad \mbox{and} \quad
^*{v}_\mu = \pm i v_\mu.
\label{dualp2}
\ee
Comparing Eqs.~(\ref{dualp1}) and (\ref{dualp2}) we see that, in order to be self dual or
anti-self dual, it is enough that the axial and vector parts of torsion be related by
\be
a_\mu = \pm \frac{2 i}{3} v_\mu.
\label{via}
\ee
Observe that the duality property (\ref{dualdual}) is preserved in this case. Observe also
that, if $v_\mu$ is assumed to be real, $a_\mu$ will be imaginary. Of course, this choice is
completely arbitrary. However, there is a physical reason to choose $v_\mu$ as real.
To understand it, let us recall that the electromagnetic coupling prescription amounts
to replace
\[
\partial_\mu \rightarrow \partial_\mu - \frac{i e}{\hbar c} A_\mu,
\]
where $A_\mu$ is the electromagnetic gauge potential. If we remember that the operator acting
on an anti-particle field is the complex conjugate, we see that the factor ``$i$'' in the
coupling constant is crucial in the sense that it is the responsible for changing the
sign of the interaction when a particle is replaced by its anti-particle. Now, differently
from electromagnetism, the gravitational interaction does not change sign when a particle is
replaced by its anti-particle. This means that the gravitational coupling prescription cannot
include the factor ``$i$'' in the coupling constant. A direct inspection of the teleparallel
Dirac equation (\ref{teledirac}) shows that $v_\mu$ must then be real, and since $\gamma^5$ is
Hermitian, $a_\mu$ must be imaginary.

It is interesting to note that, since $a_\mu$ is an axial and $v_\mu$ is a polar four-vector,
the relation (\ref{via}) presents some peculiar properties. For example, under a time reversal
operation ${\mathcal T}$,  $a_\mu$ changes sign, but not $v_\mu$. Consequently, torsion
changes from self dual (anti-self dual) to anti-self dual (self dual) under such operation. On
the other hand, under a reflection ${\mathcal P}$ of the space coordinates, $v_\mu$ changes
sign, but not $a_\mu$. Like in the previous case, torsion changes again from self dual
(anti-self dual) to anti-self dual (self dual) under such operation. Notice, however, that the
relation (\ref{via}) is invariant under a combined ${\mathcal P} {\mathcal T}$ operation.

Now, comes a crucial point: using the relation (\ref{via}), it is easy to verify that
Eqs.~(\ref{c1}) and (\ref{c2}) reduce respectively to
\be
T_{\mu \nu \lambda} \, S^{\mu \nu \rho} = - \textstyle{\frac{2}{3}} \; \delta_\lambda{}^\rho
v_\mu v^\mu
\ee
and
\be
T_{\mu \nu \theta} \, S^{\mu \nu \theta} = - \textstyle{\frac{8}{3}} \; v_\mu v^\mu.
\ee
As a simple inspection shows,
\be
T^\mu{}_{\nu \lambda} \, S_\mu{}^{\nu \rho} = 
\textstyle{\frac{1}{4}} \, \delta_\lambda{}^\rho \, T^\mu{}_{\nu \theta} \, S_\mu{}^{\nu
\theta},
\label{condualbis}
\ee
which means that, in this specific case, the condition (\ref{condual}) is fulfilled,
and the resulting gravitational theory turns out to be dual symmetric.

\subsection{Dual Symmetric Gravitation}

Let us then analyze the main properties of the above gravitational theory, which emerges from
assuming a vanishing tensorial torsion $t_{\lambda \mu \nu}$, and from the imposition that the
resulting torsion be self dual (or anti-self dual). We begin by noting that, according to
Eq.~(\ref{microtor}), torsion becomes necessarily a complex tensor. In terms of the vector
torsion, it reads
\be
T_{\lambda \mu \nu} = \textstyle{\frac{1}{3}} \left(g_{\lambda \mu} v_\nu -
g_{\lambda \nu} v_\mu \right) \pm \frac{2 i}{3} \, \epsilon_{\lambda \mu \nu \rho} \, v^\rho.
\label{dualtor2}
\ee
The corresponding contortion tensor is
\begin{equation}
K^{\rho \mu \nu} = \textstyle{\frac{1}{3}} \left(g^{\nu \rho} v^\mu -
g^{\nu \mu} v^\rho \right) \mp \frac{i}{3} \epsilon^{\nu \rho \mu \lambda} \, v_\lambda,
\label{contordual}
\end{equation}
whereas the superpotential acquires the form
\be
S^{\lambda \rho \sigma} = - \textstyle{\frac{2}{3}} \left(g^{\lambda \rho} v^\sigma -
g^{\lambda \sigma} v^\rho \right) \mp \frac{i}{3} \, \epsilon^{\lambda \rho \sigma
\theta} \, v_\theta.
\label{dualS}
\ee
The gravitational Lagrangian, however, is a real function:
\be
{\mathcal L} = - \frac{2 h}{3 k^2} \, v_\mu v^\mu.
\label{duala}
\ee
The corresponding field equation, obtained by taking its Lagrangian derivative in relation to
the gauge potential $B^a{}_\rho$, is
\be
\partial_\sigma(h S_a{}^{\rho \sigma}) = 0.
\label{bifordualbis}
\ee
As far as the theory is dual symmetric, it coincides with the Bianchi identity
(\ref{bid}) written for the dual torsion, that is, with Eq.~(\ref{bifordual}).

Now, when torsion is decomposed in irreducible parts under the global Lorentz group, the
superpotential $S_a{}^{\rho \sigma}$ becomes a complex tensor. Accordingly, the field equation
(\ref{bifordualbis}) can be split into two equations, one for the imaginary and
another for the real part. Using Eq.~(\ref{dualS}) with $S^{\lambda \rho \sigma} = h^{a
\lambda} \, S_a{}^{\rho \sigma}$, the imaginary part yields
\be
\mp \textstyle{\frac{1}{3}} \; \partial_\sigma \left( h \, h_{a \nu} \,
\epsilon^{\nu \rho \sigma \lambda} \, v_\lambda \right) = 0.
\label{eqdual1}
\ee
On account of the identity (\ref{lccoder}), it can be rewritten in the form
\be
\epsilon^{\rho \sigma \nu \lambda} \, \partial_\sigma \left( h_{a \nu} \, v_\lambda
\right) = 0,
\ee
or equivalently
\be
\partial_\sigma \left( h_{a \nu} v_\lambda \right) +
\partial_\lambda \left( h_{a \sigma} v_\nu \right) +
\partial_\nu \left( h_{a \lambda} v_\sigma \right) = 0,
\label{bidual}
\ee
which is the Bianchi identity of the theory. The real part of Eq.~(\ref{bifordualbis}), on the
other hand, is the equation governing the dynamics of the gravitational field:
\be
- \textstyle{\frac{2}{3}} \; \partial_\sigma \left[ h \left( h_a{}^\rho \, v^\sigma - 
h_a{}^\sigma \, v^\rho \right) \right] = 0.
\label{eqdual2}
\ee
In the presence of a source field, since the corresponding energy-momentum tensor
\be
\Theta_a{}^\rho \equiv - \frac{1}{h} \, \frac{\delta {\mathcal L}_M}{\delta B^a{}_\rho} =
- \frac{1}{h} \, \frac{\delta {\mathcal L}_M}{\delta h^a{}_\rho},
\ee
with ${\mathcal L}_M$ the source field Lagrangian, is always real, it contributes to the
dynamical equation only, which acquires the form
\be
- \textstyle{\frac{2}{3}} \; \partial_\sigma \left[ h \left( h_a{}^\rho \, v^\sigma - 
h_a{}^\sigma \, v^\rho \right) \right] = k^2 \, (h \Theta_a{}^\rho).
\label{eqdual2bis}
\ee

It is important to note that, since torsion is self dual (or anti-self dual), the Bianchi
identity written for {\em torsion} is mathematically the same as the Bianchi identity written
for the {\em dual torsion}. Furthermore, when torsion is decomposed according to
Eq.~(\ref{dualtor2}), both give rise, in addition to the Bianchi identity itself, the field
equation of the theory. In fact, as we have seen, whereas the Bianchi identity appears as the
imaginary part, the dynamic field equation is obtained as the real part of the
equation.\cite{footnote2}

Even though it has no meaning at the quantum level, it is instructive to obtain the
(classical) equation of motion of a spinless particle in the presence of the gravitational field
represented by the torsion tensor (\ref{dualtor2}). In the context of teleparallel gravity,
when the gravitational and inertial masses are assumed to coincide, this equation of motion
is given by\cite{wep}
\be
h^a{}_\mu \, \frac{d u_a}{ds} = T^\lambda{}_{\mu \nu} \, u_\lambda \, u^\nu,
\ee
with torsion playing the role of the gravitational force.\cite{anpe1} Using the torsion
definition (\ref{tor}), as well as the relation (\ref{rela}), this equation is easily seen to
be equivalent with the geodesic equation of general relativity. Substituting
$T^\lambda{}_{\mu \nu}$, as given by Eq.~(\ref{dualtor2}), the imaginary part of torsion
vanishes, and we get
\be
h^a{}_\mu \, \frac{d u_a}{ds} = \left( \delta_\mu{}^\lambda -
u_\mu \, u^\lambda \right) v_\lambda.
\ee
We see from this equation that the gravitational force in this case has the form of a
projector, and is clearly orthogonal to the particle's four-velocity.

\section{Final Remarks}

By generalizing the definition of the Hodge dual operator for the case of soldered bundles, and
working in the context of the teleparallel equivalent of general relativity, an analysis of the
duality symmetry in gravitation has been performed. The gauge structure of teleparallel
gravity, as well as the possibility of decomposing the torsion tensor in irreducible pieces
under the global Lorentz group, allows a much simpler and transparent approach to this problem
than it would be possible in the general relativity context. The basic conclusion that arose
from this initial analysis was that, at least in the general case, gravitation is not dual
symmetric.

However, there is a specific situation in which gravitation can be dual symmetric. This
particular situation was found by observing that the gravitational interaction of a Dirac
spinor involves only the vector and the axial parts of torsion.\cite{mospe} This fact suggests
that the purely tensorial part of torsion is irrelevant at the microscopic level, where spins
become important. Relying on this property, we have then considered a gravitational theory in
which only $v_\mu$ and $a_\mu$ are non-vanishing. Furthermore, requiring that torsion be self
dual or anti-self dual, the resulting gravitational theory was found to be dual symmetric. As
far as duality is intimately related with renormalizability,\cite{ash1} this theory may
eventually be more amenable to renormalization than teleparallel gravity or general relativity.
Accordingly, instead of general relativity (or its teleparallel equivalent), this is the
gravitational theory to be considered when dealing with the quantization problem.

When torsion is assumed to be self dual or anti-self dual, the vector and axial parts of
torsion turn out to be related by Eq.~(\ref{via}), and the Dirac equation (\ref{teledirac})
assumes the form
\begin{equation}
i \, \hbar \, \gamma^{\mu} \left[
\partial_{\mu} - \textstyle{\frac{1}{2}} \, v_{\mu} \left( 1 \mp \gamma^{5} \right)
\right]  \psi = m c \, \psi,
\label{telediracbis}
\end{equation}
with the upper (lower) sign referring to the self dual (anti-self dual) case. We see in this way
that a self dual (anti-self dual) torsion couples only to the left-hand (right-hand) component
of the spinor field. In other words, gravitation becomes a chiral interaction, a property that
may eventually have important consequences at the microscopic level. Observe that, similarly to
electrodynamics,\cite{weinberg} the field equation (\ref{telediracbis}) is invariant under a
chiral transformation
\[
\psi \rightarrow \gamma^5 \, \psi,
\]
except for a change of sign of the spinor mass term.

The possibility of decomposing torsion in irreducible pieces under the global
Lorentz group makes of teleparallel gravity a much more convenient theory than general
relativity to deal with the quantization problem. For example, it is immediate to
see from the teleparallel Dirac equation (\ref{teledirac}) that the tensorial part of torsion
does not contribute to the (microscopic) gravitational interaction of fermions,\cite{mospe} a
crucial property towards the construction of a dual symmetric gravitation. Notice in addition
that the Einstein--Hilbert Lagrangian of general relativity can be written in the teleparallel
equivalent form\cite{hs}
\be
- \frac{h}{2 k^2} \, \Rbol = \frac{h}{2 k^2} \left(\textstyle{\frac{2}{3}} \, t^{\lambda \mu
\nu} t_{\lambda \mu \nu} - \frac{2}{3} \, v^\mu v_\mu + \frac{3}{2} \, a^\mu a_\mu \right).
\ee 
When $t_{\lambda \mu \nu}$ is supposed to vanish, and the resulting torsion tensor is assumed
to be self dual (or anti-self dual), giving rise to the Lagrangian (\ref{duala}), the
corresponding general relativity Lagrangian does not have a simple form. In spite of this,
owing to their equivalence, all results---obtained here in the context of teleparallel
gravity---may also be true in the context of general relativity. Concerning this point, it is
interesting to write down the duality transformation of the general relativity spin connection
$\Abol^{abc}$ induced by the dual transformation (\ref{soldual}). This can be done by
remembering that, since the Weitzenb\"ock spin connection $A^{abc}$ vanishes, we see from
Eq.~(\ref{rela}) that, up to a zero connection,\cite{spincon}
\be
\Abol^{abc} = - K^{abc}.
\ee
Using the contortion definition, it is then easy to verify that
\begin{eqnarray}
^*{\Abol}{}^{abc} =\textstyle{\frac{1}{4}} \Big(\epsilon^{abde} \Abol_{de}{}^c +
\epsilon^{cbde} \Abol_{de}{}^a - \epsilon^{cade} \Abol_{de}{}^b \nonumber \\
+ \epsilon^{abce} \Abol_{de}{}^d - \epsilon^{abcd} \Abol_{de}{}^e \Big).
\end{eqnarray}
As before, the generalized dual definition takes into account all possible index contractions.
Of course, in the case of self dual torsion, both the spin connection $\Abol^{abc}$ and the
corresponding Riemannian curvature tensor $\Rbol^{ab}{}_{cd}$ will also be self dual.

It is worth to mention finally that, in the weak field limit of (macroscopic) teleparallel
gravity, the axial torsion is found not to contribute to the Newton potential. In fact, in
this limit, the vector and purely tensor parts of torsion combine themselves to yield the
Newton potential.\cite{pvz01} On the other hand, in the (microscopic) limit of the
gravitational interaction of fermions, it is the purely tensor part of torsion that does not
contribute. As a consequence, the dual gravitation will not exhibit a Newtonian limit. Of
course, this is not a problem as this theory is supposed to be valid only at the microscopic
level, where this limit is not required to hold.

\begin{acknowledgments}
The authors would like to thank R. Aldrovandi, V. V. Kassandrov, R. A. Mosna, Yu. N. Obukhov
and I. L. Shapiro for useful discussions. They would like also to thank CAPES, CNPq and FAPESP
for partial financial support.
\end{acknowledgments}

\end{document}